# Explainable LLM-driven Multi-dimensional Distillation for E-Commerce Relevance Learning

Gang Zhao[*]
zilong.zg@taobao.com
Taobao & Tmall Group of Alibaba
Beijing, China

Ximing Zhang[*]
zhangximing.zxm@taobao.com
Taobao & Tmall Group of Alibaba
Beijing, China

Chenji Lu
luchenji.lcj@taobao.com
Taobao & Tmall Group of Alibaba
Beijing, China

Hui Zhao
shuqian.zh@taobao.com
Taobao & Tmall Group of Alibaba
Beijing, China

Tianshu Wu
shuke.wts@taobao.com
Taobao & Tmall Group of Alibaba
Beijing, China

Pengjie Wang
pengjie.wpj@taobao.com
Taobao & Tmall Group of Alibaba
Beijing, China

Jian Xu
xiyu.xj@taobao.com
Taobao & Tmall Group of Alibaba
Beijing, China

Bo Zheng[†]
bozheng@alibaba-inc.com
Taobao & Tmall Group of Alibaba
Beijing, China

## Abstract

Effective query-item relevance modeling is pivotal for enhancing user experience and safeguarding user satisfaction in e-commerce search systems. Recently, benefiting from the vast inherent knowledge, Large Language Model (LLM) approach demonstrates strong performance and long-tail generalization ability compared with previous neural-based specialized relevance learning methods. Though promising, current LLM-based methods encounter the following inadequacies in practice: First, the massive parameters and computational demands make it difficult to be deployed online. Second, distilling LLM models to online models is a feasible direction, but the LLM relevance modeling is a black box, and its rich intrinsic knowledge is difficult to extract and apply online. To improve the interpretability of LLM and boost the performance of online relevance models via LLM, we propose an Explainable LLM-driven Multi-dimensional Distillation framework for e-commerce relevance learning, which comprises two core components: (1) An Explainable LLM for relevance modeling (ELLM-rele), which decomposes the relevance learning into intermediate steps and models relevance learning as a Chain-of-Thought (CoT) reasoning, thereby enhancing both interpretability and performance of LLM. (2) A Multi-dimensional Knowledge Distillation (MKD) architecture that transfers the knowledge of ELLM-rele to current deployable interaction-based and representation-based student models from both the relevance score distribution and CoT reasoning aspects. Through distilling the probabilistic and CoT reasoning knowledge, MKD improves both the semantic interaction and long-tail generalization abilities of student models. Extensive offline evaluations and online experiments on Taobao search ad scene demonstrate that our proposed framework significantly enhances e-commerce relevance learning performance and user experience.

## CCS Concepts

• **Information systems** → **Relevance assessment**; **Similarity measures**; **Language models**.

## Keywords

E-Commerce, Semantic Matching, Large Language Model, Knowledge Distillation

## 1 INTRODUCTION

With the widespread development of web, billions of people are now able to purchase desired items through e-commerce platforms such as Taobao[1] and Amazon[2]. Search engines play a crucial role in this landscape, enabling users to discover preferred products. Therefore, user experience is a very important part. It is indispensable to assess the semantic relevance between queries and products in order to filter out irrelevant items, thereby improving user experience and safeguarding long-term user satisfaction [3].

Due to the significant importance, text relevance learning has attracted considerable attention from researchers. Current neural-based relevance learning approaches can be mainly divided into two paradigms: *representation-based models* and *interaction-based models*. The representation-based models [16, 18, 20, 24, 29, 38, 55] focus on obtaining high-quality semantic representations for queries and items, and subsequently compute relevance scores through similarity measures or lightweight late-interaction modules. Due to the capability to pre-cache representations, these methods are currently widely applied in relevance learning for online scenarios. However, the lack of semantic interaction leads to performance deficiencies in current online methods, particularly evident in long-tail samples. The interaction-based models [7, 13, 33, 35, 47] conduct the interaction between queries and items during the semantic encoding process or incorporate complex interaction modules, enabling them to achieve robust performance. However, the larger amount of interactive computation poses significant challenges for their deployment in online e-commerce systems. The above two paradigms are essentially trade-offs between performance and efficiency toward different application scenarios. However, both methods preserve room for improvement in the era of LLM.

---

[*]Equally contribution. [†]Corresponding author.

[1]http://www.taobao.com
[2]http://www.amazon.com

Recently, Large Language Model (LLM) [45] demonstrates strong performance on relevance learning benefiting from its vast inherent knowledge. Specifically, Mehrdad et al. [27] propose to directly generate relevance judgement result for queries and items via a 7B decoder-only LLM, which achieves state-of-the-art performance and improves the generalization capabilities for long-tail samples. Due to limits of computational resources, directly deploying LLM online encounters significant challenges and is not cost-effective. Recent works [19] have shown that distilling LLM into small language models is an effective approach. However, most related works [56] only use data distillation as a means to improve efficiency. For relevance learning tasks, the decision-making process is highly interpretable and can be decomposed into query and item matching along different aspects (e.g., category, brand, etc.). Therefore, we can use the reasoning capabilities of LLM to explicitly produce such decision process, and further improve the online model performance through the intrinsic knowledge distillation, which is highly promising and explainable. Moreover, this capability is applicable to both representation-based and interaction-based models. Following the above insights, in this paper, we firstly and formally define explainable knowledge for relevance judgement along multiple aspects, and propose an *Explainable LLM-driven Multi-dimensional Knowledge Distillation framework* (**ELLM-MKD**) for e-commerce relevance learning. Our framework mainly consists of two key components:

Firstly, an **E**xplainable **LLM** for **rele**vance modeling (**ELLM-rele**) is proposed to explicitly improve the interpretability and performance of LLM and provide chain-of-thought (CoT) [53] reasoning for downstream applications. As Figure 1 shows, the key insight of ELLM-rele is to enable LLM to develop a stable CoT capability by modeling relevance judgement as a reasoning task, thus explicitly output reasoning knowledge and demonstrate the relevance assessing process. In detail, we first decompose the relevance assessment criteria into several fine-grained aspects according to the actual needs of e-commerce and annotate CoT demonstration examples by locating the relevance evidence under various aspects. Then, we perform supervised fine-tuning on a LLM via the CoT annotations, enabling it acquire stable CoT reasoning capability. With CoT reasoning, ELLM-rele can aquire better performance and interpretability by demonstrating how it makes right or fault relevance discrimination decisions.

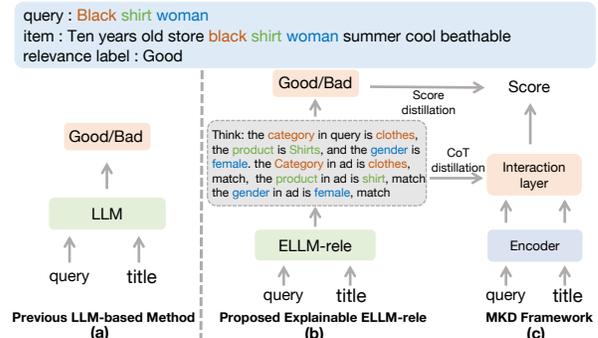

Figure 1: The key insight of ELLM-rele and MKD.

Secondly, a **M**ulti-dimensional **K**nowledge **D**istillation (**MKD**) architecture based on ELLM-rele is designed to further enhance smaller and more efficient online relevance model via maximumly exploitation of such knowledge. Specifically, MKD can distill the knowledge of ELLM-rele into both interaction and representation based models from two dimensions: *relevance score distribution* and *CoT reasoning knowledge*. From the score distribution aspect, we convert the probabilities of generated relevance judgement token within the vocabulary space into corresponding relevance scores, which enables the student model to acquire score ranking knowledge of LLM by learning the distribution of relevance scores. To deeper explore the knowledge of LLM, we inject CoT reasoning knowledge into the interaction layer of student models via an auxiliary task to enhance their semantic interaction ability. Specifically, as shown in Figure 1, the CoT output highlights the noteworthy terms from the query and item during their interaction process, as well as their relevance situation. Thus, we design two CoT distillation tasks using parsed CoT as a supervising signal for token-level semantic interactionwhere CoT distillation is applied through sequence tagging for interaction-based models and attention score regulation for representation-based models. Through score distribution and CoT knowledge distillations, MKD can effectively enhance the semantic interaction and relevance learning capabilities of existing methods.

Overall, our main contributions can be summarized as follows:

1) Based on the insight that relevance assessing criteria can be decomposed into several fine-grained aspects according to the actual needs of e-commerce, we are the first to discover the potential of explicitly modeling and exploiting the decision-making process for relevance modeling via ELLM-MKD framework.

2) To explicilty model such procedure, an Explainable LLM method for e-commerce relevance learning (ELLM-rele) is introduced, which not only provide structured intrinsic knowledge needed for downstreaming applications, but also greatly enhance the performance of LLM itself.

3) To fully exploit such intrinsic knowledge, a Multi-dimensional Knowledge Distillation (MKD) architecture is designed to transfer knowledge from ELLM-rele to current online relevance methods from both score distribution and reasoning knowledge aspects to improve online performance.

4) Extensive offline evaluation and online A/B test experiments are conducted on Taobao search advertising system, demonstrating the effectiveness of our methods by showing significant improvements in relevance learning and improving consumer experience.

## 2 RELATED WORK

### 2.1 E-Commerce Relevance Learning

E-Commerce relevance Modeling is fundamentally a text matching task, which has been extensively studied due to its pivotal role in information retrieval and search engine. Early methods such as TF-IDF [37] and BM25 [40] typically relies on manually designed statistical keyword features. Though computationally efficient, these approaches cannot capture deeper semantic relationships and are limited in handling variations such as synonyms and misspellings.

With the development of deep learning and pre-trained language models [1, 28, 45, 54], neural-based relevance learning methods have gained prominence by more effective semantic modeling and better generalizability. These methods can be broadly categorized into two paradigms: *representation-based models* and *interaction-based models*. The representation-based models, usually adopting dual-encoder architectures, focus on obtaining high-quality semantic representations for queries and items, and then subsequently compute relevance scores through similarity measures or lightweight late-interaction modules. A classic example is the Deep Structured Semantic Model (DSSM) [15], which uses separate deep fully connected networks to generate embeddings for queries and items. Subsequently, LSTM-DSSM [30] and LSTM-RNN [31] incorporate RNNs to model sequential dependencies within the text. Afterwards, The integration of pre-trained language models further advances the state of the art in e-commerce relevance learning. Sentence-BERT [39] adopts a BERT-based siamese-like architecture which encodes queries and items separately and measures the similarity of sentence representations. PolyEncoder [17] projects the query embedding into multiple spaces with learnable vectors to represent more global features. Khattab et al. [21] proposes a token-level late interaction module to further enhance the information interaction between queries and items. ReprBERT [55] distills the knowledge of interaction-based BERT into representation-based model and promotes the interaction via the intermediate interaction representation. DeepBoW [25] creates high-dimensional representations with the same dimensions as the vocabulary for queries and items, aiming at capturing detailed semantic information and better explainability. These representation-based methods allow for pre-computing embeddings, making them highly efficient for deployment in online large-scale e-commerce search engines. However, the absence of fine-grained semantic interaction results in performance limitations of current online approaches, especially pronounced in handling long-tail samples.

The interaction-based models facilitate information interaction between queries and items during the semantic encoding process or integrate sophisticated interaction modules, which enables them to achieve robust performance. Architectures such as ARC-II [14] and MatchPyramid [32] utilize convolutional neural networks (CNN) [23] to capture rich hierarchical matching patterns from the matching matrix. Match-SRNN [48] further introduces recurrent neural network [9] to model the recursive matching structure and capture long-distance dependency between the interactions. DecompAtt [34] incorporates the attention mechanism to enhance interaction and alignment between query and item tokens. Additionally, BERT [7] achieves the state-of-the-art performance by enabling deep interactions between query-item pairs during the semantic encoding process of transformer-based encoder [46]. Although interaction-based methods can achieve strong performance, the inability to pre-cache representations poses significant challenges for their deployment in online e-commerce scenes.

Recently, large language models [1, 28, 45, 54] have shown impressive performance across a range of tasks due to their extensive inherent knowledge. As an extension to interaction-based method, Mehrdad et al. [27] suggest generating relevance judgments for queries and items using a 7B decoder-only LLM, which not only delivers robust performance but also enhances the generalization for long-tail samples. Despite their promise, current LLM-based approaches have certain inadequacies in practice. Firstly, the relevance modeling process remains a black box, complicating efforts to understand the reason of LLMs' improvements or to analyze any errors in judgment. This opacity also hinders the reuse of the LLM's rich intrinsic knowledge. Secondly, their large parameter size and computational requirements pose challenges for direct online deployment. How to effectively enhance the performance of online models using LLMs is an ongoing challenge.

### 2.2 Knowledge Distillation of LLM

Knowledge distillation refers to the process of transferring knowledge from a large, complex teacher model to a smaller, more efficient student model. This technique is pivotal in addressing the computational challenges and resource limitations associated with deploying large-scale models in practical applications. Most existing approaches focus on the knowledge distillation between homogeneous LLM architectures with different amount of parameters, which can be divided into two categories. Firstly, the conventional knowledge distillation methods, aimed at enabling the student LLM to achieve performance comparable to a larger teacher LLM on various tasks. These methods include employing teacher LLM to annotate data for student LLM [11, 41, 52], directly expanding the dataset [4, 12], or fitting hidden features of student and teacher for deeper knowledge transfer [10, 44]. The second group of approaches aim at transferring the specific abilities of teacher LLM to student LLM, such as instruction follwing [36, 51], multi-turn dialogue [49], or retrieval-augmented generation [2, 26] abilities. Despite the achievements, the aforementioned methods primarily focus on the distillation of general knowledge and capabilities between LLMs, offering limited assistance for online industrial scenes like e-commerce relevance learning.

To better utilize the knowledge of LLMs, several heterogeneous distillation methods are proposed for information retrieval. Srinivasan et al. [42] proposes a two-stage framework for query rewriting which trasnfers the knowledge of LLM through annotating unlabeled data. Dai et al. [5] generates pseudo-queries for unlabeled documents based on a small number of demonstrations via LLM for dense retrieval. RankGPT [43] leverages GPT-4 to generate permutations for a group of candidate passages for the reranking task. Though gaining improvements, current data augmentation methods have not explored deeper knowledge of LLMs, such as reasoning knowledge contained in the CoT ability or fine-grained relevance ranking knowledge reflected by output probability distribution. How to enhance the performance of online heterogeneous models via LLM remains an unresolved issue.

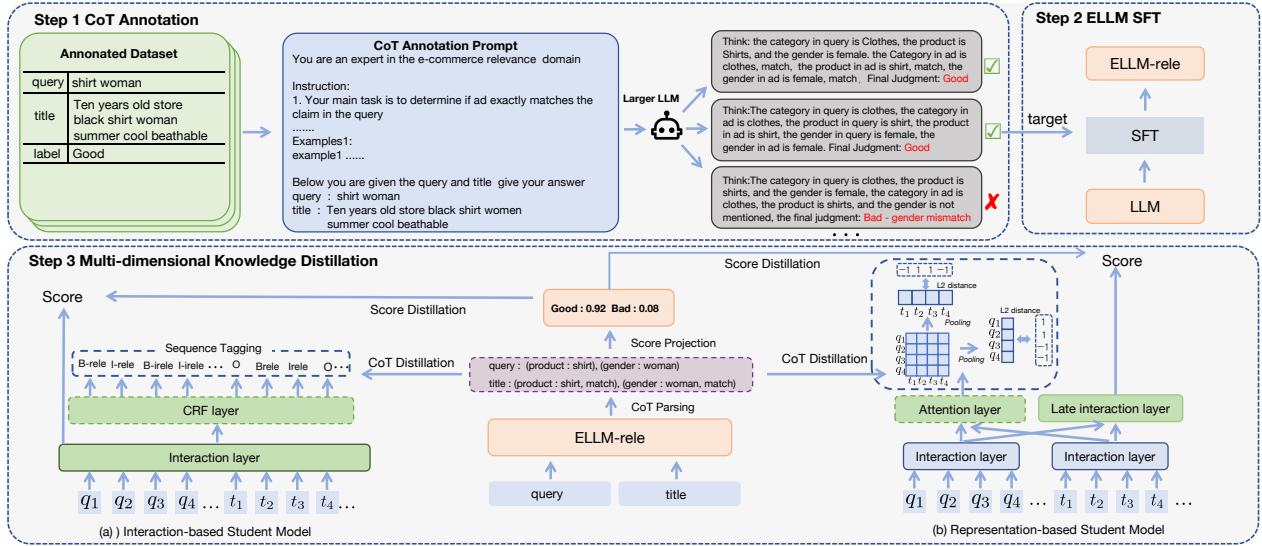

Figure 2: The overall architecture of our proposed ELLM-MKD. (1) Generating CoT annotations via Larger LLM combined with self-consistency strategy. (2) Supervised fine-tuning of ELLM-rele. (3) Multi-dimensional knowledge distillation from ELLM-rele to interaction-based and representation-based student models.

## 3 METHODOLOGY

As Figure 2 shows, the proposed framework consists of two components: Explainable LLM for relevance learning (ELLM-rele) and Multi-dimensional LLM Knowledge Distillation (MKD) module. In this section, we first introduce how we originally transform relevance learning to CoT reasoning and define multi-aspect structured CoT knowledge based on the natural application of relevance discrimination in Sec 3.1. Then, we demonstrate the detail of constructing the ELLM-rele with stable CoT reasoning ability on relevance learning in Sec 3.2. Finally, we introduce the proposed knowledge distillation approach of ELLM-rele from both the score distribution and CoT reasoning dimensions in Sec 3.3.

## 3.1 Relevance Learning as CoT Reasoning

Focused on transforming the relevance learning task to CoT reasoning for explicitly modeling and fully exploiting knowledge of LLM, we originally define multi-aspect structured CoT knowledge based on the natural application of relevance discrimination criteria. For easy understanding, we introduce some key notions here:

1) **E-commerce Relevance Learning task**: Given a user query $Q = \{q_i\}_{i=1}^{l_q}$ and an item with title $T = \{t_i\}_{i=1}^{l_t}$, where $q_i$ and $t_i$ are constituent tokens, the relevance task aims at acquiring a semantic similarity score $s(Q, T)$ indicating the query-item pair belongs to $y \in \{Good, Bad\}$ relevance situation.

2) **Chain-of-Thought** [53]: The ability of LLM to improve reasoning and decision-making by breaking down complex problems into a sequence of interconnected, logical steps, thereby enhancing the interpretability and accuracy of the model output.

3) **Relevance Learning as CoT reasoning**: To improve the interpretability and performance of LLM-based relevance method, we transform the relevance task from text classification to CoT reasoning. Given the business attributes of e-commerce scenarios, the relevance discrimination criteria can be decomposed into several fine-grained aspects such as category, brand, so the relevance task can be viewed as the matching of subproperties in the item and query. If any of the aspects do not match, we can get a fine-grained result such as "Bad - Brand mis-match", and the final judgment will be obtained as Good only if all the properties are satisfied. Based on such insight, the CoT construction of a query-item pair can be explicitly represented as: 1) attributes extracted from query, 2) attributes extracted from item title with match result to the same attribute from query, 3) final judgement as Good or Bad concluded from all attribute-level match results, as demonstrated in Figure 2. Such a design allows us to explicitly model e-commerce relevance learning as explainable CoT reasoning, not only improving the interpretability and performance of LLM-based model, but also providing additional fine-grained knowledge for further exploring. Details about the aspect definition are shown in Appendix A.

## 3.2 Explainable LLM for Relevance Learning

In this section, we introduce Explainable LLM for relevance learning (ELLM-rele) to explicitly model the relevance task as CoT reasoning, which involves two steps: high-quality CoT annotation generation and supervised fine-tuning of ELLM-rele.

*3.2.1 In-context Learning-based CoT Annotation.* We aim to leverage the CoT capacity of LLM for fine-grained extraction and discriminate relevance, where the model capacity is positively correlated with the number of model participants. In our business scenario, we need to perform daily inference on millions of data. Limited to the business requirement and computing resource, we can only choose the LLM around 7B parameters scale, which has insufficient CoT ability to meet the demands of the task without fine-tuning. Therefore, we consider generating CoT Annotations for existing labeled dataset by leveraging the few-shot capability of *larger LLM*(>70B) noted as *L-LLM* and enhance the 7B model capability by distillation. We choose two larger models Qwen2-72B[54] and LLama3-70B[8] to generate CoT Annotations, which avoids the generation of an overly homogeneous distribution of CoT data. In detail, Given the self-built dataset $D$ consisting of n triplets $\{(Q_i, T_i, L_i)\}_{i=1}^{N}$, where $Q_i$ denotes query, $T_i$ denotes item title and $L_i$ denotes label provided by the markers, we construct prompt for query-title pair as follows:

$$P_{L-LLM} = Concat(S, E, Q_i, T_i) \quad (1)$$

S denotes system prompt, which contains the task definition and the necessary instructions. E indicates few-shot examples, $Q_i$ and $T_i$ stand for the of query and item title respectively.

We prompt each model with $P_{L-LLM}$ and the self-consistency[50] strategy is utilized to increase accuracy of results. For each pair, we sampled 5 outputs from Qwen and LLama separately and grouped the results together, which can be represented as follows:

$$Collection_{CoT} = \{\arg\max P(y_i|P_{L-LLM}, \theta_{Qwen}), \\ \arg\max P(y_i|P_{L-LLM}, \theta_{Llama})\} \quad (2)$$

We finally obtain 10 CoT routes, which include the complete thought process of attribute extraction and relevance discrimination. Further, we parse the judgment result of each CoT sample and select the one whose result aligns with the corresponding label. For each query-title pair, we get a CoT annotation and the dataset $D^*$ can be constructed as $\{(Q_i, T_i, L_i, C_i)\}_{i=1}^{N}$, where $C_i$ denotes CoT annotation.

*3.2.2 Supervised Fine-tuning of ELLM-rele.* Limited by the number of parameters, the CoT capability of the original 7B LLM is not enough to accomplish the extraction and correlation discrimination tasks well. Thus we intend To enhance 7B LLM's CoT ability by data distillation based on the CoT annotation obtained from the previous Section. In detail, We take the 7B LLM as student student and the CoT annotation as its learning target. A standard SFT paradigm is employed to fine-tune 7B LLM and the negative log-likelihood loss function is optimized as follows:

$$\mathcal{L}_{LLM} = -\frac{1}{N} \sum_{i=1}^{N} \log P(C_i|S, Q_i, T_i) \quad (3)$$

In contrast to $P_{L-LLM}$, we only concatenate S, $Q_i$, and $T_i$ as the prompt $P_{LLM}$ for model fine-tuning, rather than incorporating few-shot examples. This choice is based on experimental findings that, while adding examples provides only marginal improvements to the model's performance but significantly increases inference time.

## 3.3 LLM-driven Multi-dimensional Knowledge Distillation

In this section, we detail the approach of distilling score distribution and CoT knowledge from ELLM-rele to specialist student models.

*3.3.1 Relevance Score Distribution Distillation.* Through supervised fine-tuning, ELLM-rele can learn the output template of relevance reasoning and provide a relevance decision of "*Good*" or "*Bad*" at the end of the inference sequence, which we regard as relevance judgement token. Considering that the probabilities of judgement tokens contain information about the fine-grained relevance degree, we convert them into relevance scores to guide student model training and narrow its performance gap with the teacher model.

**Data Sampling and Pseudo Label Prediction.** To sufficiently dig the knowledge of teacher model, we sample real-world unlabeled data in addition to human-annotated data for pseudo labeling of ELLM-rele. We begin by sampling query-item title pairs at the tens of millions level from Taobao online search logs, denoted as $\mathcal{D} = \{(Q_i, T_i)\}_{i=1}^{N}$. For each pair $(Q_i, T_i) \in \mathcal{D}$, the teacher model ELLM-rele generates a relevance reasoning sequence containing a judgement token $y_i \in \{\text{Good}, \text{Bad}\}$ along with the probability distribution $p_T(y|Q_i, T_i)$ over its entire vocabulary. Specifically, $p_T(y|Q_i, T_i)$ represents the likelihood of each token $y$ in the vocabulary being the output.

**Relevance Score Projection.** To isolate the relevance information from other tokens and mitigate the influence of non-relevant outputs, we focus solely on the probabilities assigned to the Good and Bad tokens. Assuming $p_T(\text{Good}|Q_i, T_i)$ and $p_T(\text{Bad}|Q_i, T_i)$ denote the probablities of relevance judgement token, we transform them into continuous relevance scores using probability normalization to ensure the scores are bounded between 0 and 1:

$$s_T(Q_i, T_i) = \frac{e^{p_T(\text{Good}|Q_i, T_i)}}{e^{p_T(\text{Good}|Q_i, T_i)} + e^{p_T(\text{Bad}|Q_i, T_i)}} \quad (4)$$

This transformation aggregates the confidence of teacher model in predicting Good versus Bad, yielding a scalar relevance score $s_T(Q_i, T_i)$ that encapsulates the semantic alignment between the query and the item title.

**Score Distillation via KL Divergence.** The distillation objective is to align the student relevance score distribution with that of the teacher ELLM-rele, we model the student score distribution $p_S(y|Q_i, T_i)$ to mimic the teacher $s_T(Q_i, T_i)$ by minimizing the Kullback-Leibler (KL) divergence between them:

$$\mathcal{L}_{\text{score}} = \frac{1}{N} \sum_{i=1}^{N} \text{KL}\left(s_T(Q_i, T_i) \,\|\, p_S(y|Q_i, T_i)\right) \quad (5)$$

By minimizing $\mathcal{L}_{\text{score}}$, the student model learns to not only align with the teacher's binary relevance decisions but also inherits the deeper probabilistic knowledge.

*3.3.2 CoT Knowledge Distillation.* To further explore the inherent knowledge of LLM, we incorporate the distillation of Chain-of-Thought (CoT) reasoning knowledge from the teacher ELLM-rele in addition to relevance score distillation. By modeling e-commerce relevance learning as an explainable CoT reasoning, ELLM-rele generates intermediate reasoning steps that elucidate the decision-making process leading to the final relevance judgment of "Good" or "Bad". These reasoning steps demonstrate fine-gained semantic interactions between the query and item title, highlighting specific relevance evidences (e.g., category, brand, etc.) appeared in query and titles as well as their alignment situation. We perform regular parsing on CoT sequences as fine-grained token-level supervision signals, and inject the interaction knowledge into the student model through auxiliary distillation tasks. Considering the scalability to adapt to both interaction-based and representation-based student paradigms, we design two kinds of CoT distillation tasks:

**CoT Distillation via Sequence Tagging.** For interaction based student models which encapsulate the query-item interaction into the semantic encoding process, such as BERT, we leverage sequence tagging on interacted embeddings to integrate CoT reasoning knowledge. Specifically, we resolve the CoT sequence to a token-level BIO (Begin, Inside, Outside) tagging scheme, where tokens corresponding to relevant evidences are annotated based on their alignment situation. Given a query $Q = \{q_i\}_{i=1}^{l_q}$ and an item $T = \{t_i\}_{i=1}^{l_t}$ where $q_i$ and $t_i$ are the constituent tokens, the CoT tagging label $\hat{C}_Q = \{\hat{c}_{q_i}\}_{i=1}^{l_q}$ and $\hat{C}_T = \{\hat{c}_{t_i}\}_{i=1}^{l_t}$ are constructed as:

$$\hat{c}_i = \begin{cases} \text{B-rele,} & \text{if } q_i/t_i \text{ is beginning of a relevant evidence} \\ \text{I-rele,} & \text{if } q_i/t_i \text{ is inside a relevant evidence span} \\ \text{B-irrele,} & \text{if } q_i/t_i \text{ is beginning of an irrelevant evidence} \\ \text{I-irrele,} & \text{if } q_i/t_i \text{ is inside an irrelevant evidence span} \\ \text{O,} & \text{if } q_i/t_i \text{ does not belong any evidence} \end{cases}$$

where $\hat{c}_i$ is the label for the $i$-th token in either the query or the item title. During training, we employ a CRF layer [22] to obtain the cot distillation loss on the interacted query and item features $\mathbf{H}_Q = \{h_{q_i}\}_{i=1}^{l_q}$ and $\mathbf{H}_T = \{h_{t_i}\}_{i=1}^{l_t}$:

$$\mathcal{L}_{\text{cot}} = -\frac{1}{N} \sum_{i=1}^{N} \left( \sum_{j=1}^{l_q} \log P(\hat{c}_{q_j}|h_{q_j}) + \sum_{j=1}^{l_t} \log P(\hat{c}_{t_j}|h_{t_j}) \right)$$

**CoT Distillation via Attention Score Regulation.** For representation based student models where the encoding and interaction processes are separated, we can improve the representation for better late-interaction by regulating the token-level attention scores based on the interaction knowledge reflected in CoT sequences. Specifically, the CoT outputs from ELLM-rele indicate relevant evidence tokens should receive higher attention scores and irrelevant ones should receive lower scores during the interaction between the query and item title. In this context, we construct an attention regulatory factor $\hat{a}_i$ for each token in the query or title as follows:

$$\hat{a}_i = \begin{cases} 1 & \text{if } q_i/t_i \text{ is a relevant evidence token} \\ -1 & \text{if } q_i/t_i \text{ is an irrelevant evidence token} \end{cases}$$

Given the encoded hidden features of the query $Q$ and the title $T$, denoted as $\mathbf{H}_Q = \{h_{q_i}\}_{i=1}^{l_q}$ and $\mathbf{H}_T = \{h_{t_i}\}_{i=1}^{l_t}$ respectively, we first perform a cross-attention mechanism to compute token-level cosine similarities, resulting in an attention matrix $\mathbf{A} = \{a_{i,j}\} \in \mathbb{R}^{l_q \times l_t}$:

$$a_{i,j} = \frac{h_{q_i} \cdot h_{t_j}}{\|h_{q_i}\| \cdot \|h_{t_j}\|}$$

where $a_{i,j}$ represents the cosine similarity between the $i$-th token in the query and the $j$-th token in the title. Subsequently, we perform max-pooling on the attention matrix $\mathbf{A}$ along both the query and title dimensions to obtain aggregated relevance scores $A_Q = \{a_{q_i}\}$ and $A_T = \{a_{t_i}\}$ for each token in the query and title, respectively:

$$a_{q_i} = \max a_{i,k}|_{k=1}^{l_t}, \quad a_{t_i} = \max a_{k,i}|_{k=1}^{l_q}$$

Then, we regulate the student attention scores based on the CoT-derived attention regulatory factors via optimizing the L2 distance for tokens with defined labels (i.e., $a_i \in \{-1, 1\}$):

$$\mathcal{L}_{\text{cot}} = -\frac{1}{N} \sum_{i=1}^{N} \left( \frac{1}{n_q} \sqrt{\sum_{\hat{a}_{q_j}} (a_{q_j} - \hat{a}_{q_j})^2} + \frac{1}{n_t} \sqrt{\sum_{\hat{a}_{t_j}} (a_{t_j} - \hat{a}_{t_j})^2} \right)$$

where $n_q$ and $n_t$ are the numbers of valid regulatory factors of query and title, and tokens which do not correspond to any evidence are excluded from the loss computation.

| Dataset | sample | query | product | Good | Bad | CoT term |
|---|---|---|---|---|---|---|
| Train | 337,066 | 84,227 | 327,303 | 244,188 | 92,878 | 1,733,110* |
| Valid | 43,887 | 31,099 | 43,595 | 32,219 | 11,668 | - |
| Test | 43,888 | 31,043 | 43,536 | 31,967 | 11,921 | - |
| Unlabeled | 30,000,000 | 6,295,824 | 11,888,256 | 23,223,531* | 6,776,469* | 119,147,259* |

**Table 1: Data statistics. * marks the values that are based on pseudo labels generated by ELLM-rele. *CoT term* refers to the number of phrases in queries and items included in the CoT.**

Finally, at the training phase, we sum the knowledge distillation losses mentioned above as well as the original cross-entropy loss as the final loss:

$$\mathcal{L} = \lambda_1 \mathcal{L}_{\text{score}} + \lambda_2 \mathcal{L}_{\text{cot}} + \frac{\lambda_3}{N} \sum_{i=1}^{N} \text{CE}(\hat{y}_i|Q_i, T_i)$$

where $\lambda_1$, $\lambda_2$ and $\lambda_3$ are hyperparameters.

## 4 EXPERIMENTS

### 4.1 Dataset

We conduct extensive experiments on large-scale human annotated e-commerce relevance dataset to validate the effectiveness of our proposed ELLM-rele and MKD architecture. The dataset contains more than 400 thousand query-item pairs sampled from the Taobao ad search logs, and then labeled *Good* (relevant) or *Bad* (irrelevant) by experienced human annotators. We divide the dataset into training, validation, and test sets for model fine-tuning and evaluation. For knowledge distillation, we collect an additional 30 million unlabeled query-item pairs randomly sampled from the Taobao search logs. The average length of queries and item titles are 6 and 34 Chinese characters. Detailed data statistics are shown in Table 1.

### 4.2 Baseline and Evaluation Metrics

*4.2.1 Baselines.* To comprehensively evaluate the effectiveness and generalizability of our proposed ELLM-MKD framework, we adopt the following diverse set of state-of-the-art baseline models across different relevance modeling paradigms. 1) *BERT* [7], an interaction-based method that jointly encodes queries and items using a bidirectional Transformer architecture to capture deep semantic interactions during the encoding process. 2) *Sentence-BERT* [38], a representation-based method which encodes queries and items separately into fixed-size embeddings for efficient computation of cosine similarity. 3) *Poly-Encoder* [18], a representation-based method which projects query embeddings into multiple vector spaces via learnable vectors to capture diverse global features. 3) *ColBERT* [20], a representation-based method that further improves the query-item interaction by employing a token-level late interaction mechanism. 4) *ReprBERT* [55], which proposes an intermediate interaction module and leverages knowledge distillation from interaction-based model to enhance semantic representations of queries and items. 5) *Warmart-LLM* [27], a LLM-based method that fine-tunes a 7B decoder-only large language model by directly generating the relevance judgments for queries and items.

*4.2.2 Evaluation Metrics.* We evaluate our method using a combination of offline and online metrics to comprehensively evaluate its performance and industrial value. For offline evaluation, given the binary nature of human annotations in the relevance task, we treat the problem as a classification task. We employ the Receiver Operating Characteristic Area Under Curve, marked as *ROC-AUC*, as a primary metric, which is commonly used in text relevance evaluations [6, 55]. Additionally, due to the imbalanced nature of e-commerce datasets where positive instances dominate, we utilize the Precision-Recall Area Under Curve. Specifically, we calculate the *Neg PR-AUC*, by treating negative instances as the positive class and positive instances as the negative class, following previous studies [24]. This metric provides a focused assessment on the model's ability to identify negative instances effectively. For online performance, we consider the proportion of "Good" annotations (*Goodrate*) as determined by human evaluators and the ad product Click-Through Rate (*CTR*). Query-item pairs for human relevance assessment are randomly sampled from online ad search logs.

| Model | ROC-AUC | PR-AUC |
|---|---|---|
| **Representation-based Model:** | | |
| Sentence-BERT♣ [38] | 78.5 | 61.3 |
| Poly-Encoder◇ [18] | 81.7 | 65.2 |
| ColBERT♡ [20] | 84.4 | 69.0 |
| ReprBERT [55] | 83.8 | 68.5 |
| **Interaction-based Model:** | | |
| BERT♠ [7] | 87.2 | 73.0 |
| **LLM-based Model:** | | |
| Walmart-LLM [27] | 91.7 | 76.4 |
| **Ours:** | | |
| ELLM-rele | **92.5** | **78.2** |
| MKD♣$_{\text{S-BERT}}$ | 81.1 | 64.6 |
| MKD◇$_{\text{Poly}}$ | 85.8 | 69.8 |
| MKD♡$_{\text{ColBERT}}$ | 87.2 | 72.9 |
| MKD♠$_{\text{BERT}}$ | 89.4 | 75.5 |

Table 2: Comparison results of different methods on the the test set ROC-AUC (%) and Neg PR-AUC (%).The best and second performances are highlighted in bold and underlined.

### 4.3 Implementation Details

*4.3.1 ELLM-rele Setup.* Considering the robust performance and adequate model size of Qwen2-7B LLM[54], we choose it as the backbone model to build ELLM-rele. We fine-tune the Qwen2-7B via the annotated CoT dataset introduced in Section 4.1. To eliminate the impact of using different backbone LLMs, the experiments of Walmart-LLM[27] are also conducted based on Qwen2-7B for a fair comparison. Both models are trained on 16 H20 GPUs with a global batch size of 256 and using AdamW optimzier to optimize the model with lr=1e-5,$\beta_1 = 0.9, \beta_2 = 0.999$,decay_rate=0.01. The models are trained for 8 thousand steps in a total of 10 hours. Besides, The self-consistency strategy is used in generating CoT annotations to improve the accuracy of Larger LLM. To balance the generation of diversity and accuracy, we set hyper-parameter TopP as 0.7, TopK as 50 and temperature T as 0.8 following previous work[50].

*4.3.2 Multi-dimentional Knowledge Distillation Setup.* To evaluate the effectiveness and generalizability of MKD on various relevance models, we perform MKD on BERT, Sentence-BERT, PolyEncoder and ColBERT baselines in offline experiments. The sequence length of queries and item titles are set to 32 and 64, the batchsize is set to 256 for all baselines and our MKD models. All models are trained via Adam optimizer with a learning rate of 5e-6 on 32 Nvidia H20 GPUs. For CoT knowledge distillation, we perform sequence taggging on BERT and attention score reguatlion on representation-based models. The MKD hyperparameters $\lambda_1, \lambda_2, \lambda_3$ are set to 1.0, 0.1, 0.5 for interaction-based model while 1.0, 0.01, 0.5 for representation-based models, respectively. For score distribution distillation, the tempreatures of KL divergence are set to 2.0. For the online experiment, our A/B test baseline model adopts the ColBERT-like architecture with only 6 transformer encoder layer and the sequence length of 16 and 32 for quries and item titles to achieve adequate computational efficiency for online services.

| Model | Params | Infer time | Train time |
|---|---|---|---|
| Sentence-BERT | 101.2M | 121 | 64663 |
| MKD$_{\text{S-BERT}}$ | 101.2M | 121 | 65400 |
| Poly-Encoder | 101.2M | 151 | 65445 |
| MKD$_{\text{Poly}}$ | 101.2M | 151 | 66182 |
| ColBERT | 101.2M | 135 | 64417 |
| MKD$_{\text{ColBERT}}$ | 101.2M | 135 | 65155 |
| BERT | 101.2M | 7344 | 54583 |
| MKD$_{\text{BERT}}$ | 101.2M | 7344 | 55210 |
| Walmart-LLM | 7B | 84000 | - |
| ELLM-rele | 7B | 432000 | - |

Table 3: Comparison of the model efficiency. *Infer/Train time* (ms) represent the corresponding averaged processing time for 1,000 query-item pairs on a single Nvidia H20 GPU.

### 4.4 Main Results

Table 2 presents the main experimental results comparing our proposed ELLM-rele and MKD against the baseline models. From the results, we can discern the following key observations:

**The superiority of LLM-based method.** Both ELLM-rele and Warmart-LLM, achieve substantially higher performance compared to expert models. Specifically, ELLM-rele surpasses the interaction-based model BERT by **5.3%** on ROC-AUC and **5.2%** on Neg PR-AUC. Additionally, ELLM-rele outperforms representation-based models such as Sentence-BERT by **14.0%** and Poly-Encoder by **10.8%** on ROC-AUC, and by **16.9%** and **13.0%** on Neg PR-AUC, respectively. These results demonstrate the efficacy of leveraging LLMs for relevance modeling, capitalizing on their extensive pre-trained knowledge to achieve significant performance enhancements.

**Effectiveness of incorporating CoT to LLM.** Specifically, ELLM-rele achieves an increase of **0.8%** in ROC-AUC and **1.8%** in Neg PR-AUC compared to Warmart-LLM. This indicates that modeling relevance as a CoT reasoning task not only enhances the model interpretability but also benefits the model performance from the fine-grained relevance discrimination and reasoning processes.

**Effectiveness and generalizability of MKD.** Furthermore, our proposed Multi-dimensional Knowledge Distillation framework achieves significant performance enhancements on student models. Specifically, MKD improve ROC-AUC by **2.6%** /**4.1%**/**2.8%**/**2.2%** and Neg PR-AUC by **3.3%**/**4.6%**/**3.9%**/**2.5%** compared to each baseline, respectively. Compared to ReprBERT, which also employs knowledge distillation, MKD can enable the already disadvantaged Poly-Encoder model to surpass ReprBERT's performance by **2.0%** on ROC-AUC, further revealing the potential of using LLM as the teacher model. The results validate the effectiveness of MKD in exploiting and transferring the knowledge from ELLM-rele to current specialized relevance models. Notably, MKD consistently delivers performance improvements across different model paradigms, underscoring its generalizability and robustness for relevance modeling in diverse settings. Further long-tail and case analysis are presented in Appendix B and C, respectively.

## 4.5 Model Complexity

Considering industrial deployment, we evaluate the model complexity and computational efficiency using parameter size and training/inference time consumption. Table 3 presents the efficiency results of various models, from which we can observe that: Firstly, the introduction of MKD does not alter the model parameter size or inference time compared to their respective baselines. This is because knowledge distillation is applied solely during the training phase, ensuring that the deployment-phase efficiency remains consistent with the baseline models. The incorporation of auxiliary distillation tasks leads to a slight increase in training time, however it remains acceptable as the training is conducted offline, without impacting the online deployment efficiency. Regarding ELLM-rele, modeling relevance learning as a CoT reasoning task leads to an increase in the length of the output tokens. This augmentation elevates inference times. Given that LLMs are predominantly deployed offline, the acceptable increase in latency is outweighed by the substantial gains in model performance and interpretability.

## 4.6 Ablation Study

To further understand the contribution of each component in our ELLM-rele and MKD framework, we perform ablation study on the following model variants: (1) *w/o CoT reasoning*, an ELLM-rele variant that removes the CoT reasoning process, modeling the relevance task as directly outputting relevance judgment tokens based on query-item pair. (2) *w/o score distillation*, a MKD variant that removes the score distribution distillation process by setting $\lambda_1 = 0$, performing only hard label distillation of ELLM-rele. (3) *w/o CoT distillation*, a MKD variant that removes the CoT distillation process by setting $\lambda_2 = 0$. (4) *w/o score&CoT*, a MKD variant that removes both the score and CoT distillation by setting $\lambda_1 = \lambda_2 = 0$. (5) *w/o MKD*, which completely removes all unlabeled data with pseudo label generated by ELLM-rele.

The ablation study results are presented in Table 4. From the results, we can observe that:

(1) Modeling the relevance judgment task as a CoT reasoning process results in an improvement of +0.8% in ROC-AUC and +1.8% in Neg PR-AUC for ELLM-rele. This demonstrates that CoT not only enhances the interpretability of the LLM-based relevance model but also contributes to performance gains by enabling finer-grained relevance discrimination and reasoning.

(2) Utilizing soft label distillation, where the language model's relevance judgment token probabilities are transformed into relevance scores, leads to an improvement of +0.9%/+0.6% in ROC-AUC and +1.6%/+0.9% in Neg PR-AUC for $\text{MKD}_{\text{ColBERT/BERT}}$. This validates the superiority of score distribution distillation over hard label distillation, as it allows the student models to capture more nuanced relevance information from the teacher model.

(3) Distilling the interaction knowledge from the CoT reasoning outputs further enhances performance by +0.5%/+0.4% in ROC-AUC and +0.9%/+0.5% in Neg PR-AUC. Furthermore, it can also bring improvements that can be overlaid with score distillation by +1.1%/+0.8% in ROC-AUC and +2.0%/+1.2% in Neg PR-AUC. This underscores the effectiveness of CoT distillation in transferring in-depth knowledge from ELLM-rele to student models.

| Model | ROC-AUC(%) | PR-AUC(%) |
|---|---|---|
| **ELLM-rele** | **92.5** | **78.2** |
| w/o CoT reasoning | -0.8 | -1.8 |
| **$\text{MKD}_{\text{ColBERT}}$** | **87.2** | **72.9** |
| w/o score distillation | -0.9 | -1.6 |
| w/o CoT distillation | -0.5 | -0.9 |
| w/o score&CoT | -1.1 | -2.0 |
| w/o MKD | -2.8 | -3.9 |
| **$\text{MKD}_{\text{BERT}}$** | **89.4** | **75.5** |
| w/o score distillation | -0.6 | -0.9 |
| w/o CoT distillation | -0.4 | -0.5 |
| w/o score&CoT | -0.8 | -1.2 |
| w/o MKD | -2.2 | -2.5 |

**Table 4: Ablation study of ELLM-rele and MKD.**

(4) Incorporating pseudo-labeled data through ELLM-rele leads to an improvement of +2.8%/+2.2% in ROC-AUC and +3.9%/+2.5% in Neg PR-AUC. This highlights the significance of utilizing large-scale unlabeled data for knowledge distillation via LLM, thereby enhancing the model's ability to generalize and accurately discriminate relevance in diverse query-item pairs.

## 4.7 Online Evaluation

We conduct online A/B test by integrating MKD into the ColBERT-like relevance model that is previously deployed, while keeping other factors unchanged. Both variations of the experiment are exposed to 5% of Taobao's search advertising traffic and run continuously for two weeks. The findings show that the introduction of MKD leads to a +0.17% increase in click-through rate (CTR). Human evaluations further indicate an overall improvement in relevance Goodrate by +0.89%, with a particularly notable increase of +1.96% for long-tail samples. These results confirm the effectiveness of the ELLM-MKD framework in enhancing relevance performance and user experience. MKD has already served the entire Taobao search advertising traffic, and ELLM-rele functions not only as the teacher relevance model but also provides additional services, such as relevance annotation assistance and error analysis.

## 5 CONCLUSION

In this paper, we propose the Explainable LLM-driven Multidimensional Distillation framework for e-commerce relevance learning, benefiting from two aspects: Firstly, to improve the interpretability and performance of current LLM-based method, we propose an Explainable LLM for relevance modeling, which decomposes the relevance learning into intermediate steps and models relevance learning as a Chain-of-Thought reasoning. Secondly, to enhance the performance of online relevance models via LLM, we propose a Multi-dimensional Knowledge Distillation architecture that transfers the knowledge of ELLM-rele to current interaction-based and representation-based student models from the relevance score distribution and CoT reasoning aspects. Extensive offline and online experimental results demonstrate the effectiveness of our proposed method in enhancing e-commerce relevance modeling performance and consumer experience.

## References

[1] Josh Achiam, Steven Adler, Sandhini Agarwal, Lama Ahmad, Ilge Akkaya, et al. 2023. Gpt-4 technical report. *arXiv preprint arXiv:2303.08774* (2023).

[2] Akari Asai, Zeqiu Wu, Yizhong Wang, Avirup Sil, and Hannaneh Hajishirzi. 2023. Self-RAG: Learning to Retrieve, Generate, and Critique through Self-Reflection. *arXiv preprint arXiv:2310.11511* (2023). https://arxiv.org/abs/2310.11511

[3] David Carmel, Elad Haramaty, Arnon Lazerson, Liane Lewin-Eytan, and Yoelle Maarek. 2020. Why do people buy seemingly irrelevant items in voice product search? On the relation between product relevance and customer satisfaction in ecommerce. In *Proceedings of the 13th international conference on web search and data mining*. 79–87.

[4] Haixing Dai, Zheng Liu, Wenxiong Liao, Xiaoke Huang, Yihan Cao, Zihao Wu, Lin Zhao, Shaochen Xu, W. Liu, Ninghao Liu, Sheng Li, Dajiang Zhu, Hongmin Cai, Lichao Sun, Quanzheng Li, Dinggang Shen, Tianming Liu, and Xiang Li. 2023. AugGPT: Leveraging ChatGPT for Text Data Augmentation. https://api.semanticscholar.org/CorpusID:257631936

[5] Zhuyun Dai, Vincent Zhao, Ji Ma, Yi Luan, Jianmo Ni, Jing Lu, Anton Bakalov, Kelvin Guu, Keith B. Hall, and Ming-Wei Chang. 2022. Promptagator: Few-shot Dense Retrieval From 8 Examples. *ArXiv* abs/2209.11755 (2022). https://api.semanticscholar.org/CorpusID:252519173

[6] Jesse Davis and Mark Goadrich. 2006. The relationship between Precision-Recall and ROC curves. In *Proceedings of the 23rd international conference on Machine learning*. 233–240.

[7] Jacob Devlin. 2018. Bert: Pre-training of deep bidirectional transformers for language understanding. *arXiv preprint arXiv:1810.04805* (2018).

[8] Abhimanyu Dubey, Abhinav Jauhri, Abhinav Pandey, Abhishek Kadian, Ahmad Al-Dahle, Aiesha Letman, Akhil Mathur, Alan Schelten, Amy Yang, Angela Fan, et al. 2024. The llama 3 herd of models. *arXiv preprint arXiv:2407.21783*.

[9] Jeffrey L. Elman. 1990. Finding Structure in Time. *Cognitive Science* 14, 2 (1990), 179–211.

[10] Yuxian Gu, Li Dong, Furu Wei, and Minlie Huang. 2024. MiniLLM: Knowledge Distillation of Large Language Models. In *The Twelfth International Conference on Learning Representations*. https://openreview.net/forum?id=5h0qf7IBZZ

[11] Xingwei He, Zheng-Wen Lin, Yeyun Gong, Alex Jin, Hang Zhang, Chen Lin, Jian Jiao, Siu Ming Yiu, Nan Duan, and Weizhu Chen. 2023. AnnoLLM: Making Large Language Models to Be Better Crowdsourced Annotators. In *North American Chapter of the Association for Computational Linguistics*. https://api.semanticscholar.org/CorpusID:257805087

[12] Zexue He, Marco Tulio Ribeiro, and Fereshte Khani. 2023. Targeted Data Generation: Finding and Fixing Model Weaknesses. In *Annual Meeting of the Association for Computational Linguistics*. https://api.semanticscholar.org/CorpusID:258960506

[13] Baotian Hu, Zhengdong Lu, Hang Li, and Qingcai Chen. 2014. Convolutional neural network architectures for matching natural language sentences. *Advances in neural information processing systems* 27 (2014).

[14] Baotian Hu, Zhengdong Lu, Hang Li, and Qingcai Chen. 2015. Convolutional Neural Network Architectures for Matching Natural Language Sentences. *Advances in neural information processing systems* 3 (2015).

[15] Po-Sen Huang, Xiaodong He, Jianfeng Gao, Li Deng, Alex Acero, and Larry Heck. 2013. Learning deep structured semantic models for web search using clickthrough data. *Proceedings of the 22nd ACM international conference on Information & Knowledge Management* (2013). https://api.semanticscholar.org/CorpusID:8384258

[16] Po-Sen Huang, Xiaodong He, Jianfeng Gao, and pages=2333–2338 year=2013 others, booktitle=Proceedings of the 22nd ACM international conference on Information & Knowledge Management. [n. d.]. Learning deep structured semantic models for web search using clickthrough data.

[17] Samuel Humeau, Kurt Shuster, Marie-Anne Lachaux, and Jason Weston. 2019. Poly-encoders: Architectures and Pre-training Strategies for Fast and Accurate Multi-sentence Scoring. In *International Conference on Learning Representations*. https://api.semanticscholar.org/CorpusID:210063976

[18] Samuel Humeau, Kurt Shuster, Marie-Anne Lachaux, and Jason Weston. 2019. Poly-encoders: Transformer architectures and pre-training strategies for fast and accurate multi-sentence scoring. *arXiv preprint arXiv:1905.01969* (2019).

[19] Martin Josifoski, Marija Sakota, Maxime Peyrard, and Robert West. 2023. Exploiting Asymmetry for Synthetic Training Data Generation: SynthIE and the Case of Information Extraction. arXiv:2303.04132 [cs.CL] https://arxiv.org/abs/2303.04132

[20] Omar Khattab and Matei Zaharia. 2020. Colbert: Efficient and effective passage search via contextualized late interaction over bert. In *Proceedings of the 43rd International ACM SIGIR conference on research and development in Information Retrieval*. 39–48.

[21] O. Khattab and Matei A. Zaharia. 2020. ColBERT: Efficient and Effective Passage Search via Contextualized Late Interaction over BERT. *Proceedings of the 43rd International ACM SIGIR Conference on Research and Development in Information Retrieval* (2020). https://api.semanticscholar.org/CorpusID:216553223

[22] John Lafferty, Andrew McCallum, Fernando Pereira, et al. 2001. Conditional random fields: Probabilistic models for segmenting and labeling sequence data. In *Icml*, Vol. 1. Williamstown, MA, 3.

[23] Y Lecun and L Bottou. 1998. Gradient-based learning applied to document recognition. *Proc. IEEE* 86, 11 (1998), 2278–2324.

[24] Zhe Lin, Jiwei Tan, Dan Ou, Xi Chen, Shaowei Yao, and Bo Zheng. 2024. Deep Bag-of-Words Model: An Efficient and Interpretable Relevance Architecture for Chinese E-Commerce. In *Proceedings of the 30th ACM SIGKDD Conference on Knowledge Discovery and Data Mining*. 5398–5408.

[25] Zhe Lin, Jiwei Tan, Dan Ou, Xi Chen, Shaowei Yao, and Bo Zheng. 2024. Deep Bag-of-Words Model: An Efficient and Interpretable Relevance Architecture for Chinese E-Commerce. In *Proceedings of the 30th ACM SIGKDD Conference on Knowledge Discovery and Data Mining* (Barcelona, Spain) *(KDD '24)*. Association for Computing Machinery, New York, NY, USA, 5398–5408. https://doi.org/10.1145/3637528.3671559

[26] Hongyin Luo, Tianhua Zhang, Yung-Sung Chuang, Yuan Gong, Yoon Kim, Xixin Wu, Helen M. Meng, and James R. Glass. 2023. Search Augmented Instruction Learning. In *The 2023 Conference on Empirical Methods in Natural Language Processing*. https://openreview.net/forum?id=noIvPGG8P1

[27] Navid Mehrdad, Hrushikesh Mohapatra, Mossaab Bagdouri, Prijith Chandran, Alessandro Magnani, Xunfan Cai, Ajit Puthenputhussery, Sachin Yadav, Tony Lee, ChengXiang Zhai, et al. 2024. Large Language Models for Relevance Judgment in Product Search. *arXiv preprint arXiv:2406.00247* (2024).

[28] Long Ouyang, Jeffrey Wu, Xu Jiang, et al. 2022. Training language models to follow instructions with human feedback. *Advances in neural information processing systems* 35 (2022), 27730–27744.

[29] Hamid Palangi, Li Deng, Yelong Shen, et al. 2014. Semantic modelling with long-short-term memory for information retrieval. *arXiv preprint arXiv:1412.6629* (2014).

[30] H. Palangi, L. Deng, Y. Shen, J. Gao, X. He, J. Chen, X. Song, and R. Ward. 2014. Semantic Modelling with Long-Short-Term Memory for Information Retrieval. *Computer Science* (2014).

[31] Hamid Palangi, Li Deng, Yelong Shen, Jianfeng Gao, Xiaodong He, Jianshu Chen, Xinying Song, and Rabab Kreidieh Ward. 2015. Deep Sentence Embedding Using Long Short-Term Memory Networks: Analysis and Application to Information Retrieval. *IEEE/ACM Transactions on Audio, Speech, and Language Processing* 24 (2015), 694–707. https://api.semanticscholar.org/CorpusID:3337266

[32] Liang Pang, Yanyan Lan, Jiafeng Guo, Jun Xu, and Xueqi Cheng. 2016. Text Matching as Image Recognition. (2016).

[33] Liang Pang, Yanyan Lan, Jiafeng Guo, Jun Xu, Shengxian Wan, and Xueqi Cheng. 2016. Text matching as image recognition. In *Proceedings of the AAAI Conference on Artificial Intelligence*, Vol. 30.

[34] Ankur Parikh, Oscar Tckstrm, Dipanjan Das, and Jakob Uszkoreit. 2016. A Decomposable Attention Model for Natural Language Inference. (2016).

[35] Ankur P Parikh, Oscar Täckström, Dipanjan Das, and Jakob Uszkoreit. 2016. A decomposable attention model for natural language inference. *arXiv preprint arXiv:1606.01933* (2016).

[36] Baolin Peng, Chunyuan Li, Pengcheng He, Michel Galley, and Jianfeng Gao. 2023. Instruction Tuning with GPT-4. *arXiv preprint arXiv:2304.03277* (2023).

[37] Juan Ramos et al. 2003. Using tf-idf to determine word relevance in document queries. In *Proceedings of the first instructional conference on machine learning*, Vol. 242. Citeseer, 29–48.

[38] N Reimers. 2019. Sentence-BERT: Sentence Embeddings using Siamese BERT-Networks. *arXiv preprint arXiv:1908.10084* (2019).

[39] Nils Reimers and Iryna Gurevych. 2019. Sentence-BERT: Sentence Embeddings using Siamese BERT-Networks. (2019).

[40] Stephen Robertson, Hugo Zaragoza, et al. 2009. The probabilistic relevance framework: BM25 and beyond. *Foundations and Trends® in Information Retrieval* 3, 4 (2009), 333–389.

[41] Kumar Shridhar, Alessandro Stolfo, and Mrinmaya Sachan. 2022. Distilling Reasoning Capabilities into Smaller Language Models. In *Annual Meeting of the Association for Computational Linguistics*. https://api.semanticscholar.org/CorpusID:258762841

[42] Krishna Srinivasan, Karthik Raman, Anupam Samanta, Ling-Yen Liao, Luca Bertelli, and Michael Bendersky. 2022. QUILL: Query Intent with Large Language Models using Retrieval Augmentation and Multi-stage Distillation. *ArXiv* abs/2210.15718 (2022). https://api.semanticscholar.org/CorpusID:253224416

[43] Weiwei Sun, Lingyong Yan, Xinyu Ma, Pengjie Ren, Dawei Yin, and Zhaochun Ren. 2023. Is ChatGPT Good at Search? Investigating Large Language Models as Re-Ranking Agent. *ArXiv* abs/2304.09542 (2023). https://api.semanticscholar.org/CorpusID:258212638

[44] Inar Timiryasov and Jean-Loup Tastet. 2023. Baby Llama: knowledge distillation from an ensemble of teachers trained on a small dataset with no performance penalty. *ArXiv* abs/2308.02019 (2023). https://api.semanticscholar.org/CorpusID:260611172

[45] Hugo Touvron, Thibaut Lavril, Gautier Izacard, Xavier Martinet, Marie-Anne Lachaux, Timothée Lacroix, Baptiste Rozière, Naman Goyal, Eric Hambro, Faisal Azhar, et al. 2023. Llama: Open and efficient foundation language models. *arXiv preprint arXiv:2302.13971* (2023).

[46] Ashish Vaswani, Noam Shazeer, Niki Parmar, Jakob Uszkoreit, Llion Jones, Aidan N Gomez, Lukasz Kaiser, and Illia Polosukhin. 2017. Attention Is All You Need. *arXiv* (2017).
[47] Shengxian Wan, Yanyan Lan, Jun Xu, Jiafeng Guo, Liang Pang, and Xueqi Cheng. 2016. Match-srnn: Modeling the recursive matching structure with spatial rnn. *arXiv preprint arXiv:1604.04378* (2016).
[48] Shengxian Wan, Yanyan Lan, Jun Xu, Jiafeng Guo, Liang Pang, and Xueqi Cheng. 2016. Match-SRNN: Modeling the Recursive Matching Structure with Spatial RNN. *arXiv e-prints* (2016).
[49] Guan Wang, Sijie Cheng, Xianyuan Zhan, Xiangang Li, Sen Song, and Yang Liu. 2024. OpenChat: Advancing Open-source Language Models with Mixed-Quality Data. In *The Twelfth International Conference on Learning Representations*. https://openreview.net/forum?id=AOJyfhWYHf
[50] Xuezhi Wang, Jason Wei, Dale Schuurmans, Quoc Le, Ed Huai hsin Chi, and Denny Zhou. 2022. Self-Consistency Improves Chain of Thought Reasoning in Language Models. *ArXiv* abs/2203.11171 (2022). https://api.semanticscholar.org/CorpusID:247595263
[51] Yizhong Wang, Yeganeh Kordi, Swaroop Mishra, Alisa Liu, Noah A. Smith, Daniel Khashabi, and Hannaneh Hajishirzi. 2022. Self-Instruct: Aligning Language Model with Self Generated Instructions.
[52] Yidong Wang, Zhuohao Yu, et al. 2023. PandaLM: An Automatic Evaluation Benchmark for LLM Instruction Tuning Optimization. *ArXiv* abs/2306.05087 (2023). https://api.semanticscholar.org/CorpusID:259108266
[53] Jason Wei, Xuezhi Wang, Dale Schuurmans, Maarten Bosma, Fei Xia, Ed Chi, Quoc V Le, Denny Zhou, et al. 2022. Chain-of-thought prompting elicits reasoning in large language models. *Advances in neural information processing systems* 35 (2022), 24824–24837.
[54] An Yang, Baosong Yang, Binyuan Hui, Bo Zheng, Bowen Yu, Chang Zhou, Chengpeng Li, Chengyuan Li, Dayiheng Liu, Fei Huang, et al. 2024. Qwen2 technical report. *arXiv preprint arXiv:2407.10671* (2024).
[55] Shaowei Yao, Jiwei Tan, Xi Chen, Juhao Zhang, Xiaoyi Zeng, and Keping Yang. 2022. ReprBERT: distilling BERT to an efficient representation-based relevance model for e-commerce. In *Proceedings of the 28th ACM SIGKDD Conference on Knowledge Discovery and Data Mining*. 4363–4371.
[56] Zhen Zhang, Yuhua Zhao, Hang Gao, and Mengting Hu. 2024. LinkNER: Linking Local Named Entity Recognition Models to Large Language Models using Uncertainty. arXiv:2402.10573 [cs.CL] https://arxiv.org/abs/2402.10573

## A  Details of CoT Construction

To better illustrate the CoT design of ELLM-rele discussed in Section 3.2.1, we provide detailed information on the aspects of e-commerce relevance discrimination in Table 5.

| Aspect | Definition |
| --- | --- |
| category | The primary categories of good, such as clothing and shoe. |
| product | Specific product, such as shirt and sneakers. |
| brand | Brand of good. |
| model number | Specific good type. |
| store | Stores that sell good. |
| color | Color of the good. |
| material | Component materials of good. |
| main Parts | Merchandise is a masterpiece or accessory. |
| elegance | The style of the product, for example, thickened. |
| season | Season of application of goods. |
| sets and singles | Product is a set or a single piece. |
| gender | Gender of the product. |
| age | Applicable age of good. |
| other properties | Other important attributes, such as second-hand. |

Table 5: Sub-dimensions of relevance discrimination for constructing the CoT of ELLM-rele.

## B  Long-tail Analysis

To evaluate the performance on queries with different occurrence frequencies, we conduct the main experiments by splitting the entire queries into head and long-tail sets by the daily user view (UV). Specifically, we define the queries with UV > 10 on the test day as head queries and the rest as long-tail queries. Note that the long-tail queries (UV <= 10) cover nearly 30% of the impressions on our online system.

| Model | Head Set | Long-tail Set | Overall |
| --- | --- | --- | --- |
| ColBERT | 84.6 | 75.5 | 84.4 |
| $MKD_{ColBERT}$ | **87.4** | **79.6** | **87.2** |

Table 6: Comparison results of ROC-AUC(%) on head and long-tail test sets.

As shown in Table 6, we find the performance of all methods decreases with the query frequency of occurrence. The performance gap between ColBERT and $MKD_{ColBERT}$ increases as the frequency decreases. Specifically, $MKD_{ColBERT}$ achieves a 4.1% ROC-AUC improvement comparing to ColBert, which shows that the performance on low-frequency queries can be improved by our multi-dimensional knowledge distillation method. In addition, the performance of Goodrate on long-tail queries in online experiments also demonstrates that the ability to distinguish long-tail queries has been significantly enhanced.

## C  Case Study

To provide qualitative insights into the performance of MKD, we present case study on typical examples. From Figure 3 we can observe that, due to the rich knowledge and CoT reasoning ability, ELLM-rele has shown good performance in cases that current student model ColBERT cannot handle well. And such knowledge can be generalized to $MKD_{ColBERT}$ through score distribution and CoT distillation, which enables the student model to make correct relevance discriminations.

**Case1**
Query : Modal pajamas for women
Title : Pajamas women 11 years old store super 20 colors cotton pajamas 2021 fall new, 100% cotton
Groud truth : Bad-material Mismatch

| Model | Output |
| --- | --- |
| ELLM | Thinking: query has category requirement:pajamas, material requirement: modal, gender requirement: female; ad in category: pajamas, matches, ad in material: cotton, mismatches, ad in gender: female, matches. material requirement in query is not met. Final judgment: Bad-material Mismatch ✅ |
| BERT | Good ✗ |
| BERT-MKD | Bad ✅ |

**Case2**
Query : Make curly hair straight
Title : Hair straightener protein correction home softener, long-lasting without damaging hair
Groud truth : Good

| Model | Output |
| --- | --- |
| ELLM | Thinking: query has category requirement:Toiletries, product requirement: make curly hair straight; ad in category: Toiletries, matches, ad in product: hair straightening cream, matches. Final judgment: Good ✅ |
| BERT | Good ✗ |
| BERT-MKD | Bad ✅ |

Figure 3: Typical translated relevance learning cases.